\documentclass[seceq,supplement]{ptptex}

\usepackage{bm, graphicx,color}
\usepackage{graphicx}
\usepackage{wrapft}

\title{
Velocity autocorrelation function of fluctuating particles in
incompressible fluids.%
}

\subtitle{Toward direct numerical simulation of particle dispersions}    

\author{
Takuya \textsc{Iwashita}$^1$, Yasuya \textsc{Nakayama}$^2$, and Ryoichi \textsc{Yamamoto}$^{1,3}$%
}

\inst{
$^1$ Department of Chemical Engineering, Kyoto University,Kyoto 615-8510,Japan\\
$^2$ Department of Chemical Engineering, Kyushu University,Fukuoka 819-0395,Japan\\
$^3$ CREST, Japan Sience and Technology Agency, Kawaguchi 332-0012, Japan
}

\abst{
Motions of fluctuating Brownian particles in an incompressible viscous 
fluid have been studied by coupled simulations of Brownian particles and
host fluid. We calculated the velocity autocorrelation functions of 
Brownian particles and compared them with the theoretical results. 
Extensive discussions have been made on the time scales for which our 
numerical model is valid.
}

\begin{document}

\maketitle

\section{Introduction}

The motions of small fluctuating particles in viscous fluids have been studied 
for a long time.
Although theoretical or numerical analysis based on the coupled motions
of the particles and the host fluid are very complicated, 
it becomes rather simple if one considers only the particles'
motions by assuming that the host fluid degree of freedom can 
be safely projected out from the entire degree of freedom
of the dispersions.
One of such models is the well-known generalized 
Langevin equation(GLE) for Brownian particles, {\it i.e.},
\begin{eqnarray}
M_i\frac{d{\bm V}_i}{dt} &=& \int_{-\infty}^{t} ds\sum_j{ \Gamma}_{ij}(t-s){\bm
 V}_j(s) +  {\bm G}_i(t),\\
\langle {\bm G}_i(t) \cdot {\bm G}_j(0)\rangle &=& 3k_BT \Gamma_{ij} (t),
\end{eqnarray}
where $M_i$ and ${\bm V}_i$ denotes the mass and the translational velocity of the
{\it i}-th particle, respectively.
$ \Gamma_{ij}(t)$ is a friction tensor, which represents the effect of
hydrodynamic interactions(HI) between {\it i}-th and {\it j}-th particles. 
${\bm G}_i$ is the random force acting on the {\it i}-th particle 
induced by thermal fluctuations of the solvent, 
$k_B$ is Boltzmann constant, and $T$ is the temperature of Brownian particles. 

For a single spherical particle ($i=1$) immersed in a infinitely large
host fluid, the analytic form of the 
time-dependent friction\cite{Landau} is known as 
\begin{equation}
\int dt \Gamma_{11}(t)\exp(-i\omega t) = \hat{\Gamma}_{11} (-i\omega) = 6\pi\eta a (1 +a\sqrt{-i\omega/\nu} - i\omega
a^2/9\nu)
\label{gamma}
\end{equation}
where $\hat{\Gamma_{11}}(-i\omega)$ is the Fourier transform of $\Gamma_{11}(t)$ and $\omega$ is the angular frequency.
The first term 
corresponds to the 
normal Stokes friction for a spherical particle of radius $a$ in a
Newtonian fluid whose viscosity is $\eta$.
The second term represents the memory effect, which is related to the
momentum diffusion in a viscous medium. 
Here the kinematic viscosity is defined as $\nu=\eta/\rho_f$ with $\rho_f$
being the density of the fluid.
The third term corresponds to 
the effect of the acceleration of the host fluid surrounding the tagged particle when the particle is accelerated through the host fluid.
Using Eq.(\ref{gamma}), the hydrodynamic GLE can be solved analytically.
The translational velocity autocorrelation function (VACF) $\langle\bm V_i(t)\cdot\bm V_i(0)\rangle/3$  then obtained is known to
exhibit the characteristic power-law relaxation for long-time region,
which is widely known as the ``hydrodynamic long-time tail''
\cite{ANA1,ANA0,ANA2,ANA}. 

For dispersions composed of many particles interacting via HI, the situation is still not
straightforward because we do not know the true analytic expression for 
the hydrodynamic friction tensor $\Gamma_{ij}(t)$.
Some approximated expressions, such as Oseen or Rotne-Prager-Yamakawa(RPY) tensor, can be obtained by introducing the Stokes approximation, however, those
expressions completely neglect the memory effect that corresponds to the
second term of Eq.(\ref{gamma}).
This means that the hydrodynamic long-time tail can not be 
reproduced correctly with Oseen or RPY tensor.

In the present study, we developed a numerical method to take into account 
the effects of hydrodynamics directly by simultaneously solving the Navier-Stokes
equation for the host fluid with the Brownian motions of 
the particles.
We first examined the VACF for a single Brownian particle and compared it 
with the analytical form mentioned above. Secondly, we examined the rotational motions 
of a single Brownian particle.
We furthermore examined the motions of Brownian particles 
in harmonic potentials to check the validity of our method.

\section{Simulation method}

Here we briefly explain the basic equations of our numerical model
since those are explained in detail elsewhere\cite{Iwa}.
A smooth profile function $0\le\phi(x,t)\le 1$ is introduced to 
define fluid ($\phi=0$) and particle ($\phi=1$) domains on 
a regular Cartesian grid.
Those two domains are separated by thin interface regions whose 
thickness is $\xi$. 
The position of the {\it i}-th particle is 
${\bm R_i}$, the translational velocity is ${\bm V_i}$, and the rotational
velocity is ${\bm \Omega_i}$.
The motion of {\it i-}th particle with mass $M_i$ and the moment of 
inertia $\bm I_i$ is governed by the following Langevin-type equations,
\begin{eqnarray}
M_i \frac{d{\bm V}_i}{dt}&=& {\bm F}^H_i
 + {\bm F}^C_i + {\bm F}^{ex}_i + {\bm G}^V_i,\ \ \
\frac{d{\bm R}_i}{dt} = {\bm V}_i,\\
\bm I_i \cdot\frac{d{\bm \Omega}_i}{dt}&=& {\bm N}^H_i + {\bm G}^\Omega_i,
\end{eqnarray}
where ${\bm F}_i^H$ and $\bm N_i^H$ are the hydrodynamic forces and
torques acting on the {\it i}-th particle due to HI, respectively. 
$\bm F^C_i$ and $\bm F^{ex}_i$ denote the direct particle-particle interaction and  external force.
${\bm G}_i^V$ and ${\bm G}_i^\Omega$ are the random force and torque 
due to thermal fluctuations defined stochastically as
\begin{align}
\langle {\bm G}_i^V\rangle&=\langle {\bm G}_i^\Omega\rangle={\bm 0},\\
\langle {\bm G}^V_i(t)\cdot {\bm G}^V_j(0)\rangle&=3k_BT\alpha^V\delta(t)\delta_{ij},\\
\langle {\bm G}^\Omega_i(t)\cdot {\bm G}^\Omega_j(0)\rangle&=3k_BT\alpha^\Omega\delta(t)\delta_{ij},
\end{align} 
where $\alpha^V$ and $\alpha^\Omega$ are parameters to control 
the temperature $T$. 

The motions of the host fluid are governed by the Navier-Stokes equation
\begin{eqnarray}
\rho_f(\partial_t {\bm v} + {\bm v}\cdot \nabla {\bm v}) &=&-\nabla p +\eta\nabla^2 {\bm v}+\rho_f\phi{\bm f_p}
\end{eqnarray}
with the incompressible condition $\nabla\cdot{\bm v}= 0$, 
where $\bm v$ and $p$ are the velocity and the pressure fields of the host
fluid, respectively,
and $\phi \bm f_p$ is the body force 
defined so that the rigidity of the particles is automatically satisfied. 
Note that ${\bm F}_i^H$ and $\bm N_i^H$ are determined from
the body force $\phi \bm f_p$ \cite{Naka,Naka1}.

\section{Results and discussion}

A single spherical particle fluctuating in a Newtonian fluid was
simulated in the absence of external forces $\bm F^{ex}_i=0$ 
as depicted in Fig.1.
We take the mesh size $\Delta$ and $\tau=\Delta^2\rho_f/\eta$ as the units
of space and time. 
Simulations have been performed with
$\eta=1$, $a=5$, and $\xi=2$ in a three-dimensional
cubic box composed of $64 \times 64 \times 64$ grid points. 
The particle and fluid densities are identically set to be unity, $\rho_p=\rho_f=1$. 
\begin{figure}[b]
  \begin{center}
   \includegraphics[width=68mm]{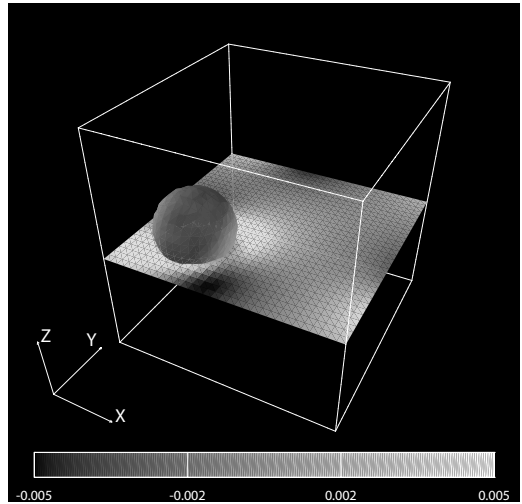}
  \end{center}
  \caption{A snapshot of a single Brownian particle immersed in a
 Newtonian fluid. 
The one eighth of the entire system is graphically displayed. 
The color map on the horizontal
plane shows the value of the local fluid velocity in the $x$ direction.}
  \label{SNAP}
\end{figure}

\begin{figure}
\begin{center}
\centerline{
\includegraphics[scale=1.]{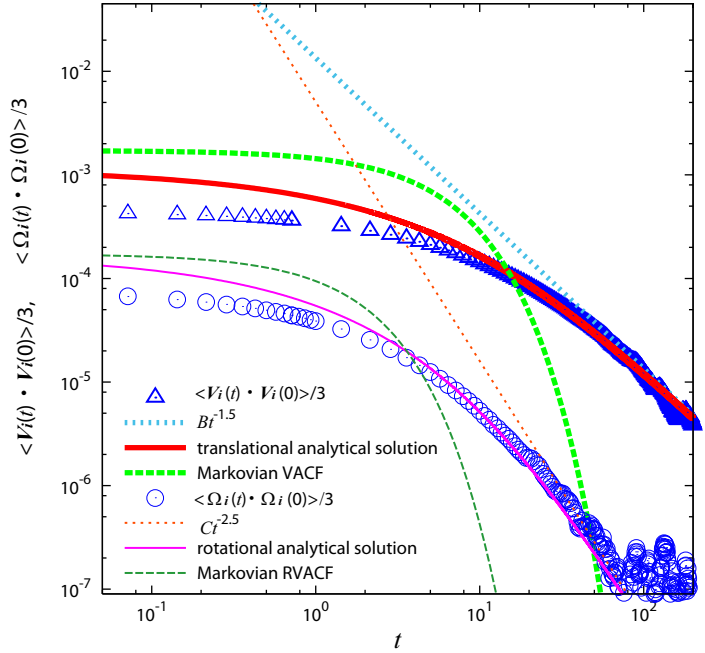}
}
\label{VACF}
\caption{The translational velocity autocorrelation function $\langle{\bm V}_i(t)\cdot{\bm V}_i(0)\rangle/3$ (triangle) and the rotational velocity autocorrelation function $\langle{\bm \Omega}_i(t)\cdot{\bm \Omega}_i(0)\rangle/3$ (circle) for a single Brownian particle 
fluctuating in a Newtonian fluid.
The simulation data was taken at $k_BT = 0.83$.
The solid lines indicate the analytic results for the
translational \cite{ANA} and the rotational motions.
The dotted lines show power-laws, 
$Bt^{-3/2}$ with $B=k_BT/12\rho_f(\pi\nu)^3$ for the translational motions 
and $Ct^{-5/2}$ with $C=\pi k_BT/32\rho_f(\pi\nu)^{5/2}$ for the rotational
motions.
The dashed lines indicate
the Markovian VACF and RVACF, which decay exponentially as $\exp(-t/\tau_B)$ and $\exp(-t/\tau_r)$, respectively.
}
\end{center}
\end{figure}

Figure 2 shows our simulation results ($\triangle$) of VACF for a single 
Brownian particle fluctuating in a Newtonian host fluid at 
$k_BT=0.83$. 
The temperature $T$ was determined by comparing the 
long-time diffusion coefficient $D_{sim}$ obtained from simulations with
$D^V=k_BT/6\pi \eta a K(\Phi)$, where $K(\Phi)$ takes into account 
the effects of finite volume fraction \cite{Zick} and $\Phi$ denotes 
the volume fraction.
The volume fraction of a single particle is $\Phi=0.002$.
One finds that the VACF approaches asymptotically to the power-law line with the exponent $-3/2$, 
and the long-time behavior of our simulation agrees well
with the analytical solution\cite{ANA} of the hydrodynamic GLE rather
than the Markovian VACF which neglects memory effects.
This behavior indicates that the memory effects are accurately taken into account.
%
Similar to the translational motions, 
we have studied the rotational motions of the Brownian particle 
in the host fluid.
The GLE of the rotational motions for a single spherical particle can be 
written as
\begin{align}
I_i\dot{\bm \Omega}_i&=-\int^t_{-\infty}ds\mu(t-s){\bm \Omega}_i(s) + {\bm G}_i(t),\\
\langle {\bm G}_i\rangle &=0, \ \ \ \langle {\bm G}_i(t) \cdot {\bm G}_i(0)\rangle = 3k_BT \mu (t),
\end{align}  
where the time-dependent friction $\mu(t)$ has the form
$\hat{\mu}(-i\omega)=8\pi\eta a^3[1 - i\omega/3\nu(1 + a\sqrt{-i\omega/\nu})]$\cite{RAMB} in Fourier space. 
The first term in $\hat{\mu}$ is the Stokes friction and the second term
represents the memory effect due to the kinematic viscosity of the
fluid. 
The GLE can be solved analytically, and the analytical solution
of the rotational velocity autocorrelation function (RVACF) is obtained
in the following form
\begin{eqnarray}
\langle {\bm \Omega}_i(t) &\cdot& {\bm \Omega}_i(0)\rangle \nonumber\\
=&& -\frac{3k_BT\nu}{8\pi\eta a^5}\int_0^{\infty}\frac{dy}{3\pi} \exp(-yt/\tau_\nu)\Biggl[ \frac{y^{3/2}}{[1-(\frac{\tau_r}{\tau_\nu} + \frac{1}{3}) y]^2 + y (1 - \frac{\tau_r}{\tau_\nu}y)^2 }\Biggl],\label{ROT_ANA}
\end{eqnarray} 
where $\tau_\nu=a^2/\nu$ and $\tau_r=I_i/8\pi\eta a^3$. 
In Fig.2, simulation results ($\bigcirc $) of RVACF are also plotted. 
%
The RVACF clearly shows the asymptotic approach to the hydrodynamic long-time tail with 
the exponent $-5/2$ which agrees well with the analytical solution 
(\ref{ROT_ANA}) rather than a simple Markovian RVACF.
By comparing the present simulation results with the corresponding 
analytical solutions more in detail, one may notice that some
discrepancies become notable for $t<\tau_B$ or $t<\tau_r$, where 
$\tau_B =M_i/6\pi\eta a=2 a^2\rho_p/9\eta\simeq5$ is the Brownian 
relaxation time and $\tau_r =I_i/8\pi\eta a^3=3\tau_B/10\simeq 1.5$ 
is the Brownian rotational relaxation time.
For opposite cases $t>\tau_B$ or $t>\tau_r$, however, 
the agreements between the numerical results and the 
analytical solutions are excellent.
This is because we neglected memory effects in thermal noises 
${\bm G}^V_i$ and ${\bm G}^{\Omega}_i$.
We however believe that the long-time behavior of our numerical model 
is valid for $t>\tau_B$ since $\tau_B$ is much longer than the memory times 
of the thermal noises.

There exist many other characteristic time-scales in particle
dispersions.
Important ones are the kinematic time-scale 
$\tau_\nu=a^2 \rho_f/\eta = 25$ which measures
the momentum diffusion over the particle size and the diffusion 
time-scale $\tau_D=a^2/D \simeq 3\times 10^3$ 
which measures the particle diffusion over the particle size.
As one can see in Fig.2, the present model works 
quite well for the time-scales comparable to $\tau_\nu$ and $\tau_D$, 
while it becomes inaccurate for $t<\tau_B$.

In order to test the validity of our method for the long-time behavior of Brownian particles, we next applied the present model to simulate Brownian particles 
fluctuating in external harmonic potentials. 
The potentials are introduced with the form 
\begin{align}
{\bm F_i}^{ex}=-k({\bm R}_i - {\bm R}_i^{eq})=-k\Delta {\bm R_i},
\end{align}
where ${\bm R}_i^{eq}$ is the $i$th particle's equilibrium position 
and $k$ is the spring constant.

Figure 3 shows the positional autocorrelation function 
$\langle\Delta{\bm R}_i(t)\cdot\Delta{\bm R}_i(0)\rangle/3$ 
of two Brownian particles in harmonic potentials whose 
minimum positions are separated by a fixed distance of $5a$. 
The pair of particles are interacting only hydrodynamically, 
and there exists no direct interactions between them.
The spring constant is set to $k=10$, and the temperature is 
$k_BT\simeq 0.0066$, which was determined by 
the average potential energy $k_BT=k\langle\Delta{\bm R}_i^2\rangle/3$. 
The simulation results ($\bigcirc $) agree well with the hydrodynamic 
analytical solution\cite{ANA} in harmonic potentials which account for the effects of finite volume fraction. 
The analytical solution was derived by solving the GLE of a single Brownian particle in a harmonic potential which includes the modified Stokes friction $\zeta=6\pi\eta a K(\Phi)$.
The correlation functions decay much slower than the Markovian
relaxation functions. 
We also confirmed that the validity of our method is excellent 
for $t\ge\tau_\nu$.
\begin{center}
\begin{figure}
\centerline{
\includegraphics[scale=1.1]{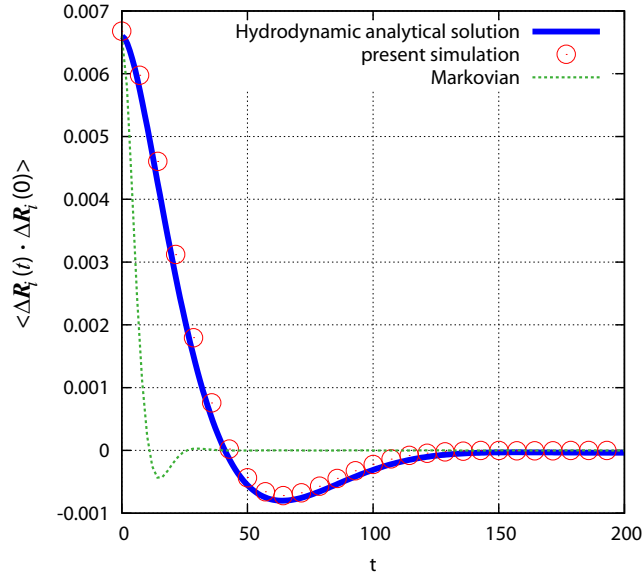}
}
\caption{The positional autocorrelation function $\langle\Delta{\bm
 R}_i(t)\cdot\Delta{\bm R}_i(0)\rangle/3$ (circle) for a system composed
 of two Brownian particles in harmonic potentials $-k({\bm R}_i - {\bm
 R}_i^{eq})$. The distance between their equilibrium positions is $|{\bm
 R}_1^{eq}-{\bm R}_2^{eq}|=5a$. The solid line shows the analytical
 solution\cite{ANA}.
The dotted line shows the Markovian functions, {\it i.e.}
the solution of 
$M_i\dot{\bm V}_i = -6\pi\eta a\bm V_i - k\bm R_i + \bm G_i.$}
\end{figure}
\end{center}
\section{Conclusion}
We proposed a numerical model to simulate Brownian particles
fluctuating in Newtonian host fluids.
To test the validity of the model, 
the translational velocity autocorrelation function (VACF), 
the rotational velocity autocorrelation function (RVACF), 
and the positional autocorrelation function of fluctuating Brownian
particles were calculated in some simple situations for which analytical
solutions were obtained.
We compared our numerical results with the analytical solutions
and found excellent agreements between them specially for long-time regions 
$t>\tau_B$ while some discrepancies were found for short time regions 
$t<\tau_B$. 
This is because our model is designed to simulate correct long-time 
behaviors of Brownian particles in host fluids.
Applications of the present method for more complicated situations 
are in progress.

\section*{Acknowledgements}
This work was partly supported by a grant from Hosokawa powder technology foundation.


%


\begin{thebibliography}{99}
\bibitem{Landau}
E.~M. Lifshitz and L.~D. Landau:
{\em Fluid Mechanics,} (Addison-Wesley, Reading, 1959).
\bibitem{ANA1}
B.~J. Alder and T.~E. Wainwright: Phys. Rev. A. {\bf 1} (1970), 18
\bibitem{ANA0}
A. Widom: Phys. Rev. A. {\bf 3} (1971), 1394
\bibitem{ANA2} E. H. Hauge and A. Martin-L$\ddot{\text{o}}$f: J. Stat. Phys. {\bf 7} (1973), 259 
\bibitem{ANA}
H.~J. H. Clercx and P.~P. J. Schram: Phys. Rev. A. {\bf 46} (1992), 1942.
\bibitem{Iwa}
T.Iwashita, Y.Nakayama, and R.Yamamoto: J. Phys. Soc. Jpn. {\bf 77} (2008), 074007.
\bibitem{Naka}
Y. Nakayama and R. Yamamoto: Phys. Rev. E. {\bf 71} (2005), 036707.
\bibitem{Naka1}
Y. Nakayama, K. Kim, and R. Yamamoto: Eur. Phys. J. E. (2008), 361.
\bibitem{Zick}
A.~A. Zick and G.~M. Homsy: J. Fluid Mech. {\bf 115} (1982), 13.
\bibitem{RAMB}
S. H. Lamb: {\em Hydrodynamics,} (DOVER PUBLICATIONS, NEW YORK, 1932).


\end{thebibliography}

\end{document}